\documentclass[
  reprint,
  amsmath,amssymb,
  aps
]{revtex4-2}

\usepackage[english]{babel}
\usepackage{ulem}
\makeatletter
\@namedef{l@en}{\l@english}
\makeatother

\usepackage[dvipsnames]{xcolor}
\usepackage{graphicx}
\usepackage{dcolumn}
\usepackage{bm}

\begin{document}

\title{\bf D-MODD: A Diffusion Model of Opinion Dynamics Derived from Online Data}
\thanks{ixandra.achitouv@cnrs.fr}%

\author{Ixandra Achitouv}
 \affiliation{Sorbonne Université, CNRS, LIP6, F-75005 Paris, France}
\affiliation{Complex Systems Institute of Paris île-de-France (ISC-PIF, UAR3611), Paris, France}
 
\author{David Chavalarias}%
\affiliation{Centre d'Analyse et de Mathématique Sociales (CAMS, UMR8557), Paris, France}%
\affiliation{Complex Systems Institute of Paris île-de-France (ISC-PIF, UAR3611), Paris, France}%

\author{Raphaël Fournier-S'niehotta}
 \affiliation{Sorbonne Université, CNRS, LIP6, F-75005 Paris, France}
\affiliation{Cédric Lab, CNAM Paris, France}
 

\begin{abstract}
We present the first empirical derivation of a continuous time stochastic model for real world opinion dynamics. Using longitudinal social media data to infer users opinion on a binary climate change topic, we reconstruct the underlying drift and diffusion functions governing individual opinion updates. We show that the observed dynamics are well described by a Langevin type stochastic differential equation, with persistent attractor basins and spatially sensitive drift and diffusion terms. The empirically inferred one-step transition probabilities closely reproduce the transition kernel generated from the D-MODD model we introduce. Our results provide the first direct evidence that online opinion dynamics on a polarized topic admit a Markovian description at the operator level, with empirically reconstructed transition kernels accurately reproduced by a data-driven Langevin model, bridging sociophysics, behavioral data, and complex-systems modeling.
\end{abstract}

\maketitle



\section*{Introduction}

Online social networks constitute a primary arena of opinion formation, where robust collective patterns such as consensus, polarization, and echo chambers emerge from repeated local interactions, despite the complexity of human psychology and algorithmic mediation ~\cite{peralta2022opiniondynamicssocialnetworks,Sasahara_2020}. Opinion dynamics models have proven powerful theoretical tools to elucidate the mechanisms driving those phenomena. Classical Models such as the DeGroot framework ~\citep{degroot_reaching_1974}, the Friedkin-Johnsen model ~\citep{friedkin_social_1990}, bounded confidence models ~\citep{deffuant_mixing_2000,hegselmann2002}, and physics inspired voter and Ising models ~\citep{holley_ergodic_1975,ising_beitrag_1925} offer theoretical mechanisms for consensus formation, polarization, and fragmentation.

Although these models have significantly deepened our understanding of opinion dynamics, few have been calibrated using time series of real world behavior  ~\cite{Liu_2023,Sasahara_2020, chavalarias_can_2024, Vendeville_2025}. Furthermore, most models rely on agent based simulations, which, although they succeed in reproducing certain emergent collective behaviors, fail to identify the stochastic laws that govern them. There remains a need to link microscopic behaviors to macroscopic outcomes through an analytical conceptualization.

In the age of large-scale social platforms, abundant behavioral traces are available, yet tracking how individual opinions evolve over time remains challenging. Most empirical studies rely on survey snapshots or aggregate sentiment measures, which provide static or coarse-grained views of public opinion ~\cite{douven_mis2021,Vendeville_2025}. Even when digital traces are available, connecting them to a continuous opinion variable suitable for dynamical analysis is non-trivial. Prior work has typically inferred polarization from network structure, clustering, or sentiment classification ~\cite{Conover2011,Barbera2015,Bail2018} but has not recovered user level opinion trajectories with sufficient temporal resolution to identify underlying dynamical laws.

In this article, we address these gaps by empirically modeling the stochastic dynamics of opinion change from empirical data. Our results can be summarized in two key results: 
(i) We measure, for the first time in a large online social network, the conditional probability that a user updates their opinion given their previous state and (ii) we use this empirical transition kernel to calibrate and propose a new opinion-dynamics model D-MODD that links observed behavior to an effective stochastic differential equation for opinion dynamics. 

This article is organized as follows. In Sec.\ref{sec1}, we present the dataset, the construction of the latent opinion space, and the validation of the inferred opinions. In Sec.\ref{sec2}, we introduce the stochastic description of opinion dynamics, including the empirical reconstruction of drift and diffusion functions and the D-MODD model. In Sec.\ref{sec3}, we analyze user profiles in relation to the inferred diffusion properties. Finally, Sec.\ref{sec4} concludes with a discussion of the implications and perspectives of our approach.


\begin{figure*}[t]
\centering
\begin{minipage}{0.32\textwidth}
\centering
\includegraphics[width=\linewidth]{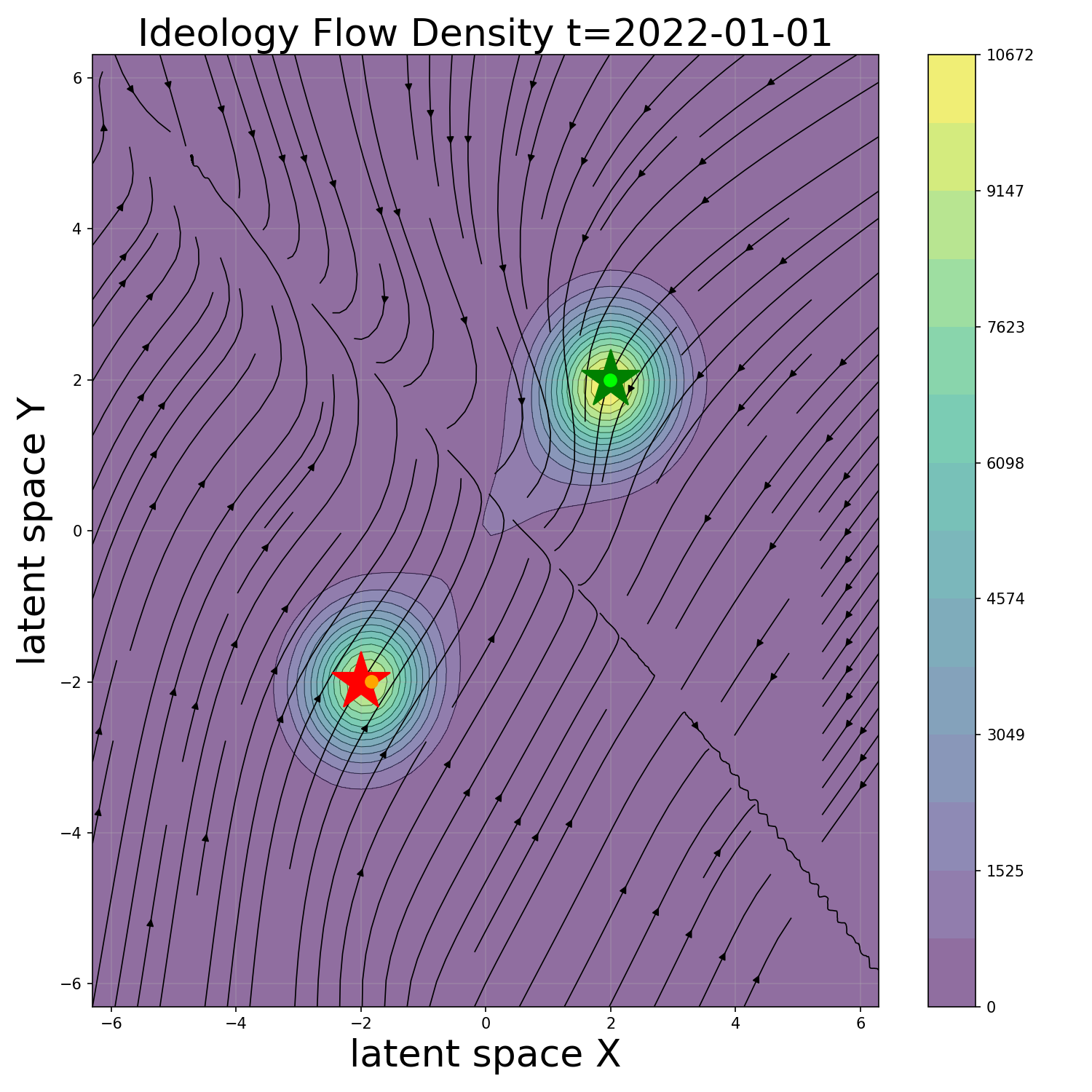}
\end{minipage}\hfill
\begin{minipage}{0.32\textwidth}
\centering
\includegraphics[width=\linewidth]{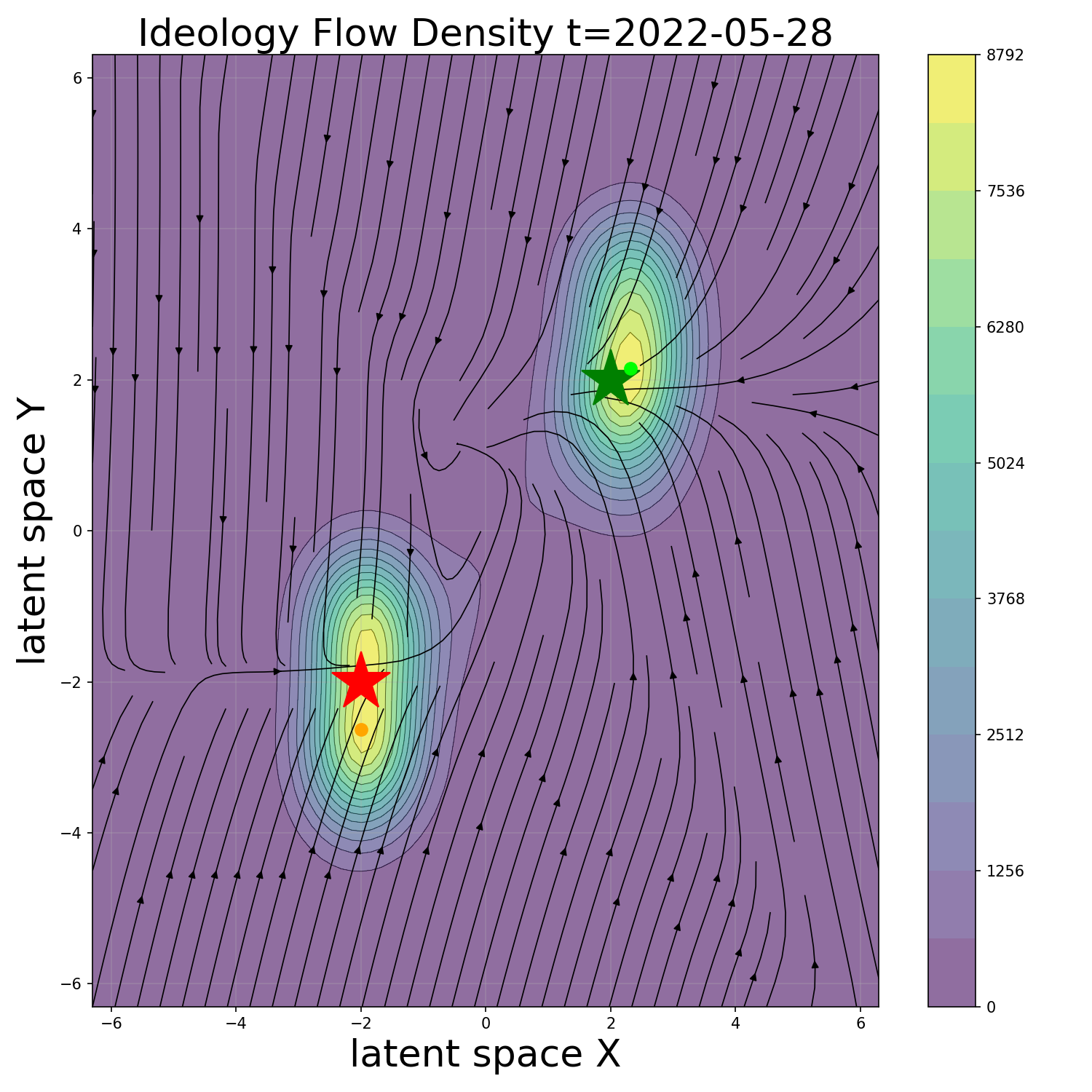}
\end{minipage}\hfill
\begin{minipage}{0.32\textwidth}
\centering
\includegraphics[width=\linewidth]{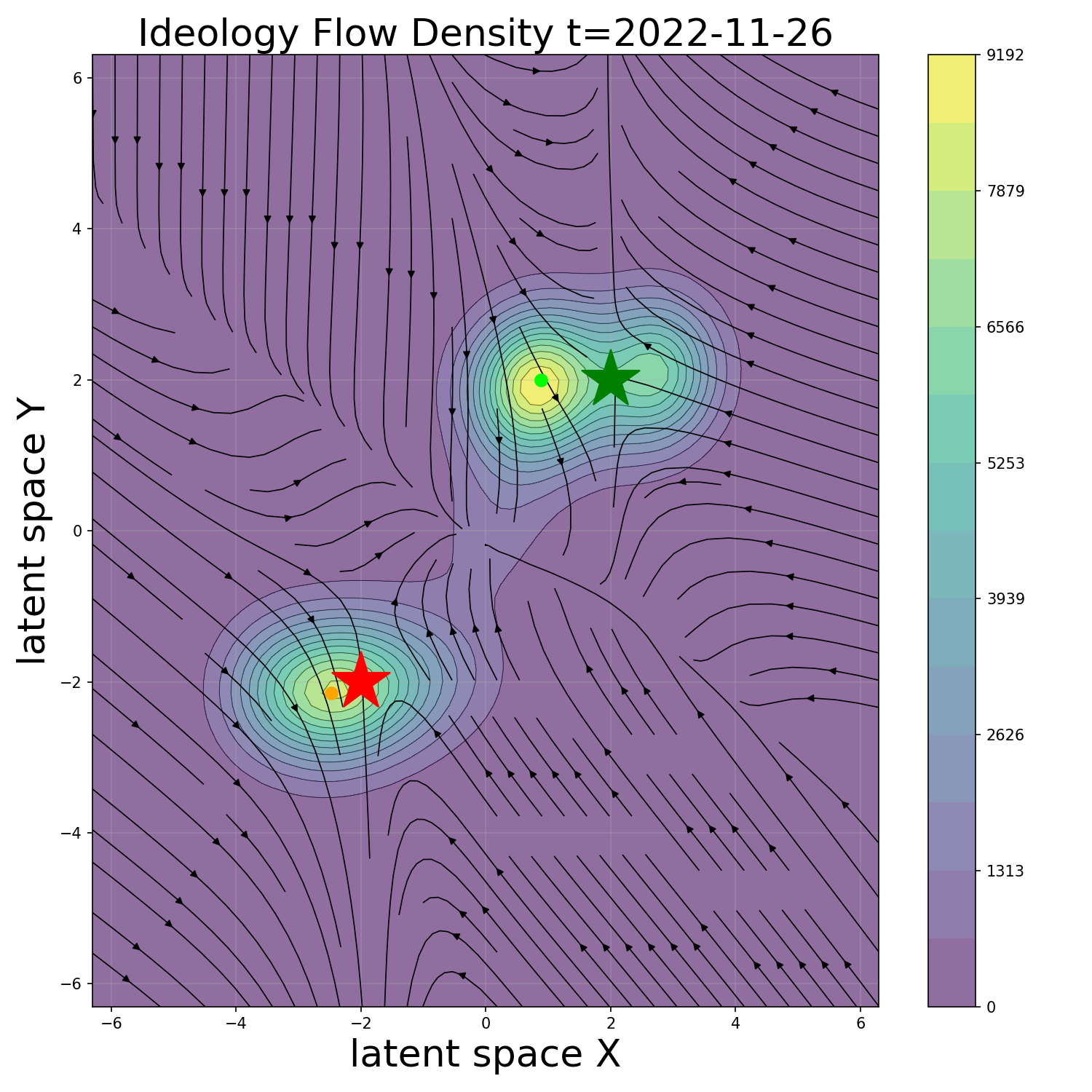}
\end{minipage}

\caption{Opinion flow density in the latent space $\mathcal{O}$ at three representative times. Green and red stars mark the barycenters of the pro-climate and denialist anchors, respectively. Colored dots indicate the locations of maximum user density.}
\label{fig:ideology_flow}
\end{figure*}

\section{ Dataset \& Methodology}\label{sec1}

\subsection{Dataset}
The Climatoscope project~\cite{chavalarias_hashtags_2025} used Twitter’s track API to collect all tweets mentioning given expressions, and based on a list of several dozens of English and French keywords related to climate change, revealing a strongly polarized opinion landscape between climate skeptics/denialist and pro-climate. Over the year 2022, this collection corresponds to 57M tweets, 32.1M of them being retweets. We draw on this highly detailed longitudinal dataset to track changes in individual opinions on climate change. First, we compute graphs of online interactions over several periods of time. These dynamical graphs are computed from retweet networks built over a three-weeks period T, with a sliding windows $dT=7$ days resulting in 50 graphs for the year 2022. The weight of an edge between two accounts equals the number of retweets. When computing these graphs, we require a threshold for the minimum number of retweets corresponding to the median distribution over T. In our case this threshold is $>1$, resulting in 691,993 unique users.

We then embed these graphs using node2vec ~\citep{grover2016node2vec} and we apply a supervised UMAP 2D reduction ~\citep{mcinnes2018umap} trained on two sets of anchor accounts with known pro-climate or denialist positions. At each time step, we compute the barycenter of the pro and denialist anchors and perform a linear coordinate transformation to require that the pro and denialist barycenters are located at the coordinates (+2, +2) and (-2, -2), respectively. This yields a two-dimensional latent space $\mathcal{O}$ in which these two sets of coordinates form an effective potential landscape for online opinion evolution and where each user account $i$ has a coordinate $(x_i,y_i)$. Indeed, the smoothed density of users concentrates around these two barycenters, while the coordinate difference between two consecutive time steps exhibits coherent flows converging toward these points, revealing two stable attractor basins with diffusive fluctuations around them. 
These two distinct and persistent attractor regions are shown in Fig.~\ref{fig:ideology_flow} where the green/red stars correspond, respectively, to the barycenters of pro-climate and denialist anchor accounts. Most users are distributed around these attractors, as we can observed from the iso-density contours and the density peaks (green/red dots, next to/overlapping with the stars). The flow field, reconstructed by Gaussian-kernel smoothing of empirical one-step opinion displacements on a regular grid, reveals that the users’ trajectories between consecutive time steps tend to converge toward these attractors. This indicates that the majority of users align with the dominant orientation of their respective communities (pro-climate vs denialist), with fluctuations whose magnitude can be quantified to extract sociologically meaningful measures of within-group opinion variability and the strength of collective alignment. For instance, the variance of the iso-contours could reflects the in-group opinion diversity of each social group.

\subsection{Inferring opinion}
\begin{figure}[h]
    \centering
    \includegraphics[width=1\linewidth]{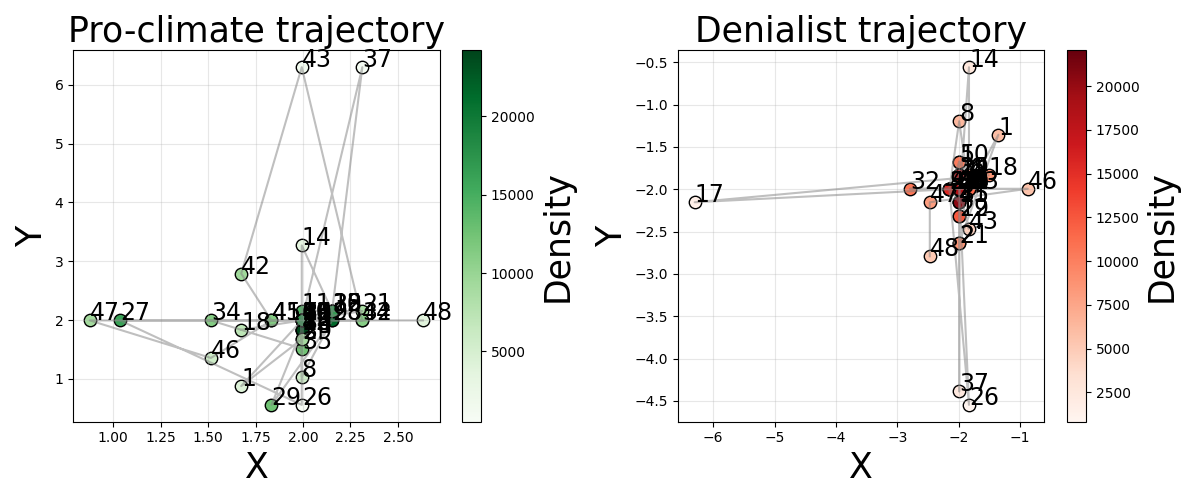}
    \caption{Trajectories of the maximum of the density in each quadrant.}
    \label{fig:traj_dens}
\end{figure}

To quantify the evolution of these attractors over time, we track the location of the density maximum within each quadrant of the latent space (Fig.~\ref{fig:traj_dens}). The quadrant $(x>0, y>0)$ corresponds to pro-climate users, while $(x<0, y<0)$ corresponds to denialist users. The positions of the density maxima fluctuate around the attractors ($\pm$2,$\pm$2) but occasionally exhibit abrupt excursions. These excursions correspond to transient drops in the density peak defining the attractor, consistent with short-lived noise-driven perturbations of an otherwise stable attractor basin. Such intermittent excursions are a hallmark of noise-driven nonlinear dynamics in bistable potential landscapes, where stochastic perturbations induce temporary destabilization of metastable attractors ~\cite{hanggi1990,freidlin2012}.

To track the users' opinion in time, we focus on the first latent coordinate $x\in \mathcal{O}_x$, which captures the dominant axis of separation between the two communities and exhibits the most stable temporal structure (smaller variance). This scalar variable defines a continuous opinion coordinate $x(t)$ for each user, allowing a high-resolution measurement of opinion updates over time depending on the minimum distance between the two attractors (effectively this means {assigning} pro-climate {opinion} for $x>0$ and denialist {ones} for $x<0$).

\subsection{Validation of the Inferred Opinions}

\begin{figure}[h]
    \centering
    \includegraphics[width=1\linewidth]{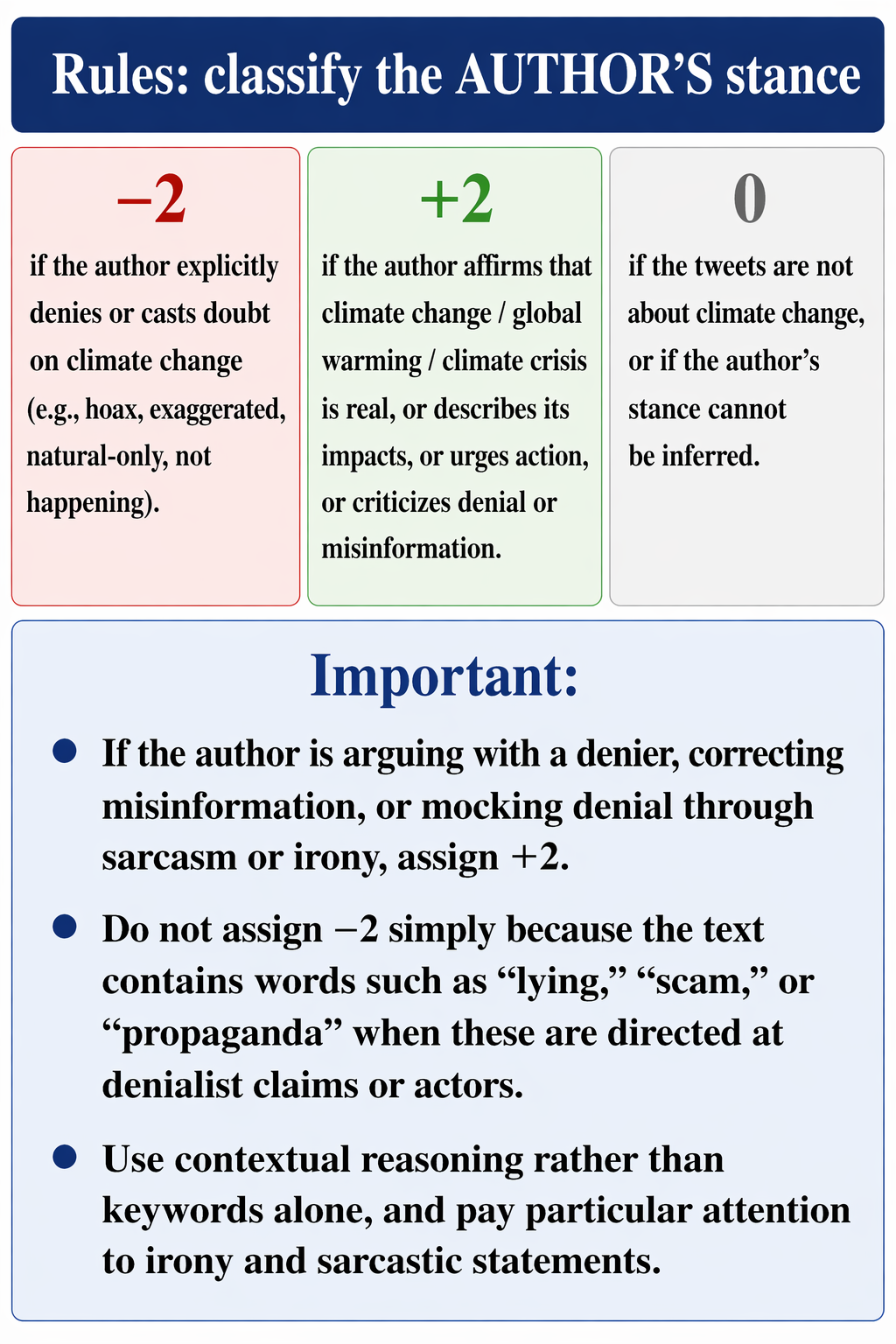}
   \caption{llm prompt to classify the stance of users}
   \label{fig:Prompt}
\end{figure}
In order to validate our methodology, we perform a number of consistency check. First we verify that the dynamics is similar for different time splits $T=\{7,14,21\}$, rolling windows $dT=\{7,14\}$ and the threshold of the number of tweets $\{1,2\}$. 

Embedding the nodes with graph neural networks such as GraphSAGE ~\cite{hamilton2017inductive} has been
attempted in both unsupervised and semi-supervised settings. While node2vec relies on biased random walks combined with a skip-gram model to learn the node embeddings in a transductive setting, GraphSAGE adopts an inductive framework that generates embeddings by aggregating features from a node’s local neighborhood.
In a similar fashion as the work on Node2Vec, GNN-based embeddings were fed to a UMAP supervised process, but the results were less accurate than those obtained with the pure random-walk method: the lack of non structural features did not help the GNNs outperform topology-based algorithm.

Third, we validate inferred opinions with two different methods. On the one hand, we compare the classification of approximately 30,100 accounts, each with at least 20 retweets and a well-defined opinion, with the classification reported in Ref.~\cite{chavalarias_hashtags_2025}, based on clustering of the retweet graph over the period from 2022-07-01 to 2022-10-30. A total of 27,700 accounts are common to both classifications. Among these, 99\% of the 17,341 accounts with $x>1$ are categorized as pro-climate in Ref.~\cite{chavalarias_hashtags_2025}, while 97\% of the 12,751 accounts with $x<-1$ are categorized as denialist pointing to a high agreement with the opinion coordinate $x(t)$

On the other hand, we also compare our classification against an independent text-based stance classification obtained from original tweets posted by randomly selected users, using the open-source \texttt{Llama-3.1} large language model~\cite{touvron2024llama3}. For this second validation, we consider only original tweets and exclude retweets, so that the textual stance estimation remains fully independent of the interaction graphs used to infer opinions. 

More precisely, we randomly select users who posted at least six original tweets during a randomly chosen time step within the year (we do not consider more than 6 tweets per user for this test). For each selected user, we aggregate the corresponding tweets and ask the language model to infer the author's stance on climate change using the prompt in Fig.\ref{fig:Prompt}

We then consider an equal number of users classified by the language model as $+2$ and $-2$, and retain only those for which at least five tweets are consistently assigned a stance of $+2$ or $-2$ (corresponding to more than $70\%$ of their tweets in that time period).

We compare this text-based classification with the opinion inferred from the dynamical embedding at the same time step. To do so, we round the inferred opinion $x(t)$ to the closest integer  and select users with $x(t)=\pm 2$. We find an agreement of $96\%$ precision over a sample of 576 users, providing strong quantitative support for the validity of the opinion inference from the dynamical embedding.

\section{Stochastic Description of Opinion Dynamics}\label{sec2}

\subsection{Empirical functions}

To characterize the opinion dynamic through its local transition statistics we compute the increment $\Delta x = x_{t+\Delta t} - x_t$ for every paired update $(x_t, x_{t+\Delta t})$  and estimate the first two conditional moments across users:
\begin{align}
    A(x) &= \frac{1}{\Delta t}\,\mathbb{E}\!\left[\Delta x \mid x_t = x\right], \\
    D(x) &= \frac{1}{2\Delta t}\,\mathbb{V}\!\left[\Delta x \mid x_t = x\right]
\end{align}
In the continuum limit, these empirical functions correspond to the drift and
diffusion coefficients of a one-dimensional Fokker--Planck operator that governs
the evolution of the opinion density $p(x,t)$,
\begin{equation}
\frac{\partial p}{\partial t}
  = -\frac{\partial}{\partial x}\!\left[A(x)\,p(x,t)\right]
    + \frac{\partial^2}{\partial x^2}\!\left[D(x)\,p(x,t)\right].
\label{eq:FP-opinion}
\end{equation}
The drift $A(x)$ captures the systematic component of opinion change, while the diffusion $D(x)$ quantifies the state-dependent variability of belief updates. For systems with multiplicative noise, drift and diffusion jointly define an effective potential landscape that shapes the stability of opinions. For one-dimensional Markovian dynamics with state-dependent diffusion, assuming a vanishing probability current, the stationary solution of Eq.\ref{eq:FP-opinion} can be written as $p_s(x) \propto e^{-V_{\mathrm{eff}}(x)}$, where the effective potential is given by
\begin{equation}
V_{\mathrm{eff}}(x)
= -\int^x \frac{A(y)}{D(y)}\,dy + \ln D(x).
\end{equation}
Thus, “opinion stability” becomes a balance between deterministic forces (restoring drift) and stochastic spreading.

\begin{figure}
    \centering
    \includegraphics[width=1\linewidth]{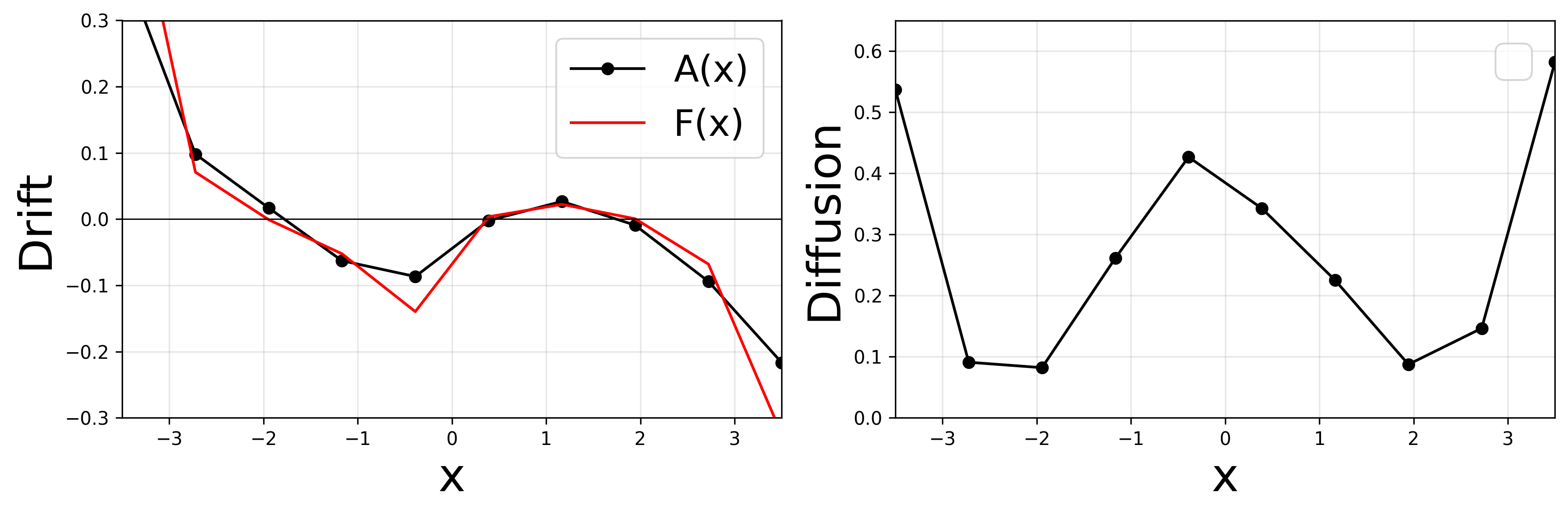}
    \caption{Drift $A(x)$, Eq.\ref{eqFx} (left) and diffusion $D(x)$ (right) estimated from
    longitudinal opinion trajectories $x$ of users.}
    \label{fig:drift_diffusion}
\end{figure}

Figure~\ref{fig:drift_diffusion} shows the empirical drift $A(x)$ and diffusion $D(x)$ coefficients reconstructed from user-level opinion trajectories (black curves).
The drift exhibits a clear restoring structure across the opinion axis: it changes sign near $x \simeq \pm 2$, corresponding to the empirically observed barycenters of the two polarized communities, and drives opinions back toward
these regions when they deviate from them. In addition, the drift crosses zero near $x \simeq 0$, identifying an unstable fixed point that separates the two basins of attraction. The restoring drift is not perfectly symmetric across the opinion axis.
Moderate asymmetries are observed in the slope and curvature of $A(x)$ around the two attractors, suggesting differences in the local geometry of the effective potential experienced by pro-climate ($x>0$) and denialist ($x<0$) users. Although these asymmetries do not indicate distinct dynamical regimes, they point to differences in the strength or spatial extent of the stabilizing forces acting on the two communities. The diffusion coefficient $D(x)$ also exhibits pronounced spatial structure. Opinion fluctuations are smallest in the vicinity of the attractor regions, but remain finite even at the stable points, reflecting persistent heterogeneity in individual-level updates. Away from the attractors, $D(x)$ increases markedly, indicating a state-dependent amplification of noise.
Finally, this multiplicative-noise regime, in which stochastic variability is strongest precisely in regions where the deterministic restoring forces weaken, is a hallmark of nonlinear stochastic systems evolving in metastable landscapes~\cite{hanggi1990}. In our case, this reveals an effective bistable dynamics with two stable attractors and a repulsive intermediate state, consistent with an interpretation in terms of a polarized opinion  ~\cite{chavalarias_hashtags_2025}. While recent theoretical approaches, such as ~\cite{Liu_2023}, explain polarization in terms of stationary bimodal opinion distributions emerging from coupled opinion–interaction dynamics, our approach complements these results by reconstructing the effective stochastic operator governing individual opinion trajectories directly from data.

\subsection{A macroscopic model of opinion dynamic: D-MODD}
Empirical drift $A(x)$ and diffusion $D(x)$ suggest that, on Twitter, the main component of opinion regarding attitudes toward climate change can be effectively described by a one-dimensional stochastic process with stable attractors and state-dependent noise. We formalize this observation in a data-driven Langevin model that we call {D-MODD} (Diffusion Model of Opinion Dynamics Derived from Data). The model provides a continuous-time generative description of how opinions evolve, calibrated directly from behavioral trajectories. At the individual level, each user's latent opinion $x(t)$ evolves according to the stochastic differential equation
\begin{equation}
    dx_t = F(x_t)\,dt + \sqrt{2D(x_t)}\,dW_t,
\label{eq:D-MODD-SDE}
\end{equation}
where $F(x)$ is the effective drift field and $D(x)$ the state-dependent
diffusion, and $W_t$ a standard Wiener process. The drift field encodes the
deterministic tendency for opinions to move toward community-specific
attractors, while the diffusion term captures the heterogeneity and
asymmetry of stochastic fluctuations observed in user-level updates. Accordingly, the deterministic drift is
modeled as
\begin{equation}
    F(x) = |A(x)|\,\bigl(x^\star - x\bigr),\label{eqFx}
\end{equation}
where $x^\star$ denotes the location of the attractor (+2 for pro-climate
users and $-2$ for denialist users).

This formulation preserves the empirically
observed restoring forces with a confining force ensuring global stability. The multiplicative factor $\bigl(x^\star - x\bigr)$ provides a
minimal stabilizing mechanism outside the empirical range, preventing numerical
divergence when simulating trajectories beyond the region where
$A(x)$ was directly observed. Eq.~\ref{eqFx} is shown in Fig.~3 (red curve) together with the empirically reconstructed drift $A(x)$ (black curve), illustrating that the regularized expression reproduces the main qualitative features of the empirical drift over the observed domain and provides a well-behaved continuation outside it. 
Simulating Eq.~(\ref{eq:D-MODD-SDE}) using an Euler--Maruyama scheme with the empirical fields $A(x)$ and $D(x)$, we produce synthetic opinion trajectories.

\begin{figure*}[t]
    \centering
    \includegraphics[width=0.95\textwidth]{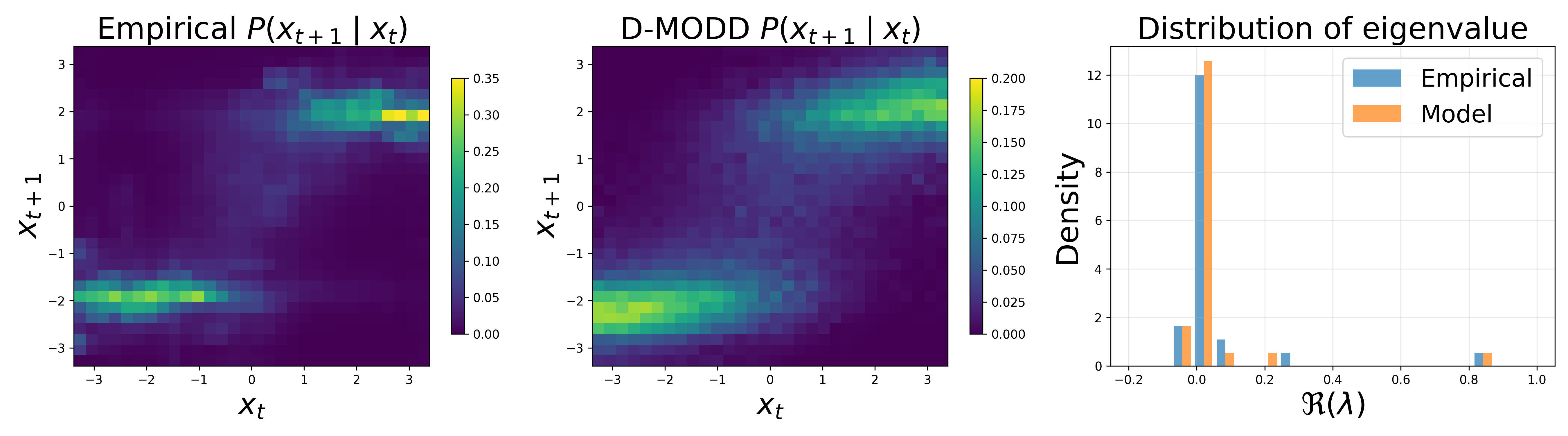}
    \caption{
    Empirical and modeled one-step opinion transition kernels.
    Left: Empirical conditional transition probability
    $P(x_{t+\Delta t}\mid x_t)$ reconstructed from longitudinal user trajectories
    in the latent opinion space.
    Middle: Transition density generated by the fitted D-MODD Langevin model
    using the empirically inferred drift $A(x)$ and diffusion $D(x)$.
    Right: Distribution of the real parts of the eigenvalues of the empirical
    and model transition operators; the Wasserstein spectral distance is $<0.01$.
    }
    \label{fig:transition_comparison}
\end{figure*}

\medskip

\subsection{Transition Kernels and Operator Validation}

A distinctive contribution of our work is that we obtain, for the first time, a full empirical reconstruction of the one-step conditional opinion transition probability $P(x_{t+\Delta t} \mid x_t)$, directly measured from longitudinal behavioral trajectories in a large online social network. Previous empirical studies of online opinion dynamics typically rely on aggregate polarization metrics, community structure, or sentiment-based ideological scores ~\cite{Conover2011,Barbera2015,Bail2018}, but do not recover a user-level transition kernel that can be used to calibrate and falsify continuous-time dynamical opinion models ~\cite{Bail2021}.

Beyond reconstructing the transition kernel, we assess whether online opinion evolution admits a Markovian description by testing the Chapman–Kolmogorov (CK) consistency condition. Empirically, we compute both the one-step and two-steps kernel $
P(x_{t+\Delta t} \mid x_t); 
P(x_{t+2\Delta t} \mid x_t) $.
Then, we construct the CK-predicted two-step kernel

\begin{multline}
P_{\mathrm{CK}}(x_{t+2\Delta t} \mid x_t)
= \\
\int P(x_{t+2\Delta t} \mid x_{t+\Delta t}) \,
     P(x_{t+\Delta t} \mid x_t) \,
     dx_{t+\Delta t} ,
\end{multline}

and we compare this to the empirical two-step kernel. We obtain that the CK-predicted kernel closely matches the empirical two-steps kernel at short times, with an average Wasserstein-1 deviation
\(
W_1 = \sum_i \pi_i\, W_1\!\left(P_{\mathrm{emp}}(i,\cdot), P_{\mathrm{CK}}(i,\cdot)\right)
\)
corresponding to $0.7$ bins of the coarse-grained state space, with $\pi_i$ the empirical distribution of visited states, and a maximum entry-wise discrepancy 
$\max_{i,j}\!\left|P_{\mathrm{emp}}(i,j)-P_{\mathrm{CK}}(i,j)\right| < 0.12$,
consistent with a first-order Markov description of short-timescale opinion evolution.

A key test of the model is therefore to compare the empirical transition kernel 
\(
P(x_{t+\Delta t} \mid x_t)
\)
with the transition density generated by the fitted Langevin model.  
Fig. \ref{fig:transition_comparison} compares the empirical one-step transition kernel reconstructed from behavioral trajectories with the transition kernel generated by the D–MODD Langevin model calibrated on measured drift $A(x)$ and diffusion $D(x)$.
The first two panels show that the model reproduces the dominant geometric features of the empirical kernel, corresponding to the two stable opinion attractors detected in the latent space including the location, width, and curvature of the probability mass around the two attractors. 

The third panel presents the distribution of the eigenvalues of the empirical and model transition operators. This spectral comparison provides a more stringent test of dynamical similarity: the eigenvalue distribution encodes the relaxation timescales, stability structure, and mixing behavior of the underlying Markov process. The close overlap between the empirical and model spectra indicates that D--MODD not only matches the point-wise transition probabilities, but also faithfully captures the global dynamical structure governing opinion evolution. Quantitatively, after excluding the trivial stationary eigenmode (\(\lambda \simeq 1\)), the spectral agreement is high, with a Wasserstein distance \(W_1 = 7.2 \times 10^{-3}\) between the empirical and model eigenvalue distributions.

\section{Attractors characterization}\label{sec3}
\begin{figure}
    \centering
    \includegraphics[width=1\linewidth]{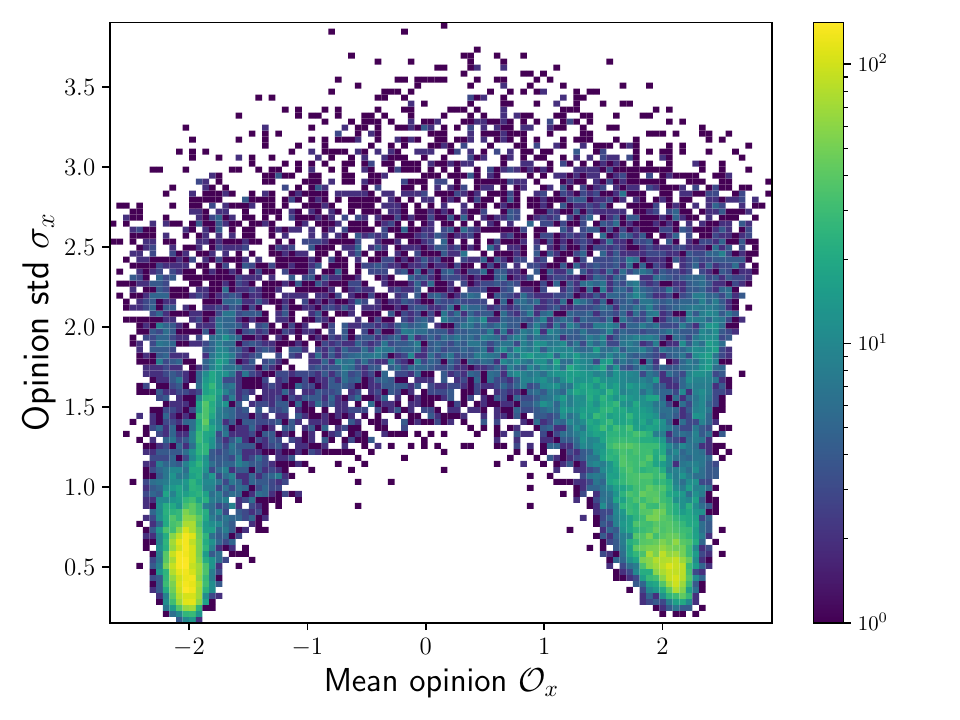}
    \caption{2D histogram of account’s opinion component $\mathcal{O}_x$ and their variance $\sigma_x$. Only users with at least 20 tweets or retweets and a clear opinion (i.e. outside the range $[-1,1]$) have been plotted. The color indicates the density of users.}
    \label{fig:hist2D}
\end{figure}

We can now analyze the fine structure of the opinion space in light of our model. First, we observe that the two attractors host different populations with respect to the relationship between opinion $x$ and opinion variance $\sigma_x$ (Fig.~\ref{fig:hist2D}). In what follows, we restrict our analysis to accounts with clear opinions, defined by $(x \in \mathcal{O}_x)$ with $\mathcal{O}_x = \{x \in \mathbb{R} : |x|>1\}$, and with at least 20 retweets. Among these 30,092 accounts, the average individual standard deviation in opinion is $\langle \sigma_x \rangle = 1.08$, while the dispersion of individual variability across the population is $\sigma_{\sigma_x} = 0.64$. 
The 17,341 accounts in the pro-climate attractor ($x>1$, representing $57.62\%$ of the sample) have a mean opinion $\langle x^{\rm pro\mbox{-}clim} \rangle = 1.86$ and an average individual variability $\langle \sigma_x^{\rm pro\mbox{-}clim} \rangle = 1.21$, which is $0.20$ above the population average. In contrast, the 12,751 denialist accounts ($x<-1$) have a mean opinion $\langle x^{\rm den} \rangle = -1.95$ and $\langle \sigma_x^{\rm den} \rangle = 0.89$, which is $0.29$ below the population average. 
These results indicate that denialist users are, on average, both more extreme in their opinions and associated with lower variability, suggesting more stable opinion trajectories.

To interpret this asymmetry, we conducted a targeted qualitative inspection of accounts with the highest and lowest $\sigma_x$ (30 in each group).

For each of these accounts, we performed the following:
\begin{itemize}
    \item Recorded their profile description
    \item Screened their last 10 to 50 tweets and manually assigned an account category (e.g., journalist, activist, NGO).
    \item Searched for tweets mentioning climate change using the query \texttt{from:screenname climate} in X's search engine. When successful, we recorded one of the climate tweets as an example. The default X search engine being limited to recent weeks, and some searches returned no results.
    \item Identified their country of origin using X's dedicated feature on the profile's homepage.
    \item Compared the climate orientation assigned by our automated method with their actual orientation.
    \item Launched a Grok profile analysis (Grok is X's AI, \url{https://x.ai}). This analysis was used only as supplementary information and treated with caution since LLMs are not fully reliable, even for summarization.
\end{itemize}


\subsection{Users with the Lowest Standard Deviation in Opinion}

\begin{figure}
    \centering
    \includegraphics[width=1\linewidth]{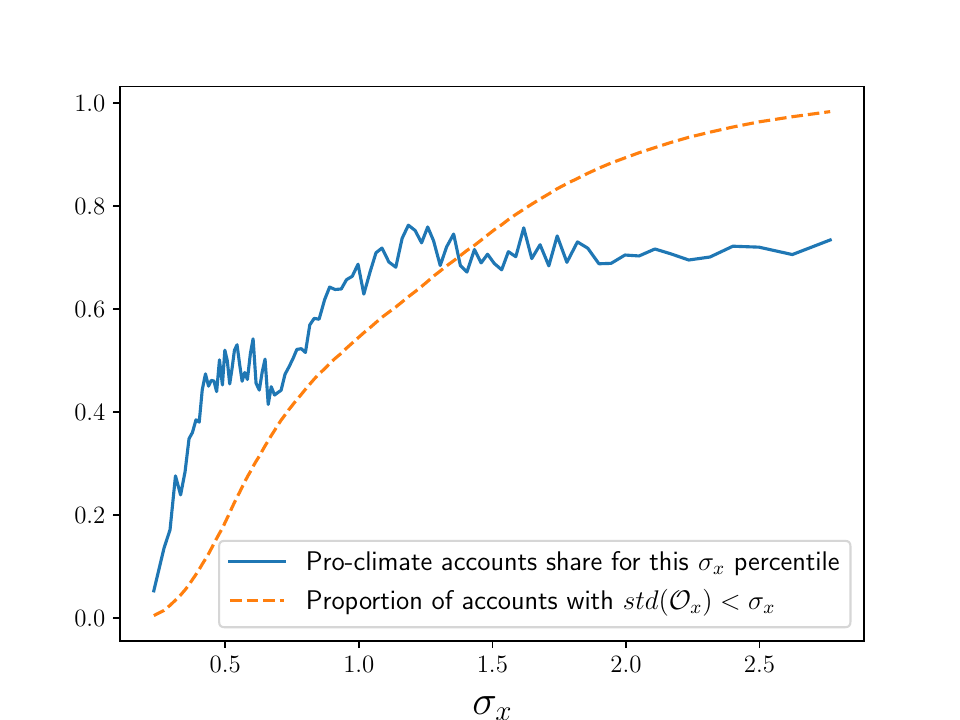}
    \caption{Empirical share of pro-climate users as a function of $\sigma_x$. The set of individual standard deviations $\{\sigma_x^k\}_{k\in \mathcal{U}}$ is partitioned into 100 bins $b_i = [\sigma_x^{(i)}, \sigma_x^{(i+1)}]$, with median value $\sigma_m^i$. Each bin defines a subset of users $\mathcal{U}_i = \{k \in \mathcal{U} \mid \sigma_x^k \in b_i\}$. The blue curve shows the fraction of pro-climate users in each bin, defined as $|\mathcal{U}_i^{\mathrm{pro}}| / |\mathcal{U}_i|$, where $\mathcal{U}_i^{\mathrm{pro}} = \{k \in \mathcal{U}_i \mid x_k > 1\}$, plotted at $\sigma_m^i$. The dashed curve represents the cumulative distribution function $F(\sigma) = |\{k \in \mathcal{U} \mid \sigma_x^k \leq \sigma\}| / |\mathcal{U}|$.}
    \label{fig:MeanOp}
\end{figure}

The average opinion of the first percentile of users with the lowest $\sigma_x$ ($\sigma_x \in [0.149, 0.25]$) is $-1.78$, with only $5.3\%$ of these 300 users categorized as pro-climate (Fig.~\ref{fig:MeanOp}).

The 30 users with the lowest $\sigma_x$ are all classified as climate deniers by the automated method, with an average opinion of $-2$. At the time of manual inspection, five of these accounts had been deleted and one was protected. Among the remaining 24 accessible accounts, all were manually identified as climate deniers, yielding full agreement with our automated classification on this subset. Of these 24 accounts, 19 are based in the United States (86.3\%), two have no identified country, and the remaining three are located in Belgium, the Netherlands, and Brazil.

While climate-related posts could not be retrieved for nine of these 24 accounts, 8 out these 9 could still be classified as aligned with MAGA ideology, which suggest climate-skeptic attitudes. Overall, all accounts for which we could screen the timeline identified themselves as MAGA supporters, conservatives, or libertarians. None were classified as pro-climate. Moreover, all those 30 accounts were from individuals identified by terms such “husband”, “wife”, “mother”, “vibe coder”, “business owner”.

The climate deniers attractor thus comprises the most hardline accounts, which represent individuals primarily from the United States who subscribe to the MAGA ideology.

\subsection{Users with the Highest Standard Deviation in Opinion}

A contrasting pattern is observed for users with high $\sigma_x$. Among the 100 users with the highest $\sigma_x$ ($\sigma_x \in [2.88, 3.66]$), only 16 were initially classified as climate deniers by the automated method. Manual inspection indicates that five of these were misclassified, while no clear opinion could be assigned to five additional accounts. In addition, one account initially classified as pro-climate was identified as a denialist account exhibiting inconsistent or disruptive posting behavior. Note that a higher rate of misclassification is expected in this regime, as users with large $\sigma_x$ exhibit more heterogeneous or context-dependent activity, making their stance more difficult to assign reliably.

After manual screening, only 6\% of the 100 accounts with the highest $\sigma_x$ were categorized as climate deniers, with an average opinion of $1.28$. Among the 30 accounts with the highest $\sigma_x$ ([3.34, 3.66]), three were protected, three had been deleted, and climate-related tweets could not be retrieved for three others. Two accounts were environmental advocates, one was neutral, three were skeptics, and the rest were categorized as pro-climate. Overall, among the 21 accounts for which climate-related posts could be retrieved, only three (14.2\%) were climate deniers, while 16 (76\%) actively supported climate measures.

Regarding countries of origin, only 3 (10\%) of the 30 users with the highest $\sigma_x$ were from the U.S. The others came from a wide range of countries and continents, with the UK being the most represented (5 accounts). Their types of accounts was diversified, one half being institutional of professional accounts (6 NGO, 2 UN accounts, 1 government account,  3 accounts from journalists, 2 from climate scientists, 1 from a company). The other half were personal accounts and activist accounts. In contrast with low $\sigma_x$ users, the high $\sigma_x$ observed in some accounts can be interpreted as a consequence of their institutional nature: since these accounts primarily disseminate factual or corporate information, their content may cover a wide range of topics and be shared by both pro-climate and climate-skeptic audiences.

In conclusion, this manual analysis confirms that the more stable of the two attractors is that of the climate deniers, characterized by less volatility in the opinions of its core accounts (lower $\sigma_x$). Opinion dynamics literature has highlighted the role of stubborn or extreme agents in emergence of polarization and the stability of opinion attractors ~\cite{galam_contrarian_2004, perrier_phase_2024, amblard_role_2004} showing that only a tiny fraction of stubborn agents in a population could change the attractors of the opinion landscape, from convergence to consensus to polarization.

In the case of the climate change debate, the presence of "stubborn" denialists some of whom are supporters of the MAGA movement could perpetuate a divide among part of the population regarding what is scientifically recognized as a consensus on the human-caused origins of climate change. We were unable to determine whether these core accounts are genuine or fake. 

\section{ Discussion and Conclusions}\label{sec4}

In this article, we have developed a data-driven stochastic description of opinion dynamics by empirically reconstructing the transition operator governing individual opinion updates in a large online social network.

The Chapman-Kolmogorov validation and the spectral analysis demonstrates that real-world online opinion fluctuations can be accurately described by a low-dimensional stochastic process whose drift and diffusion functions are directly inferred from behavioral data. By reconstructing the full conditional transition kernel with hundred thousands of empirical opinion updates, we provide the first data-driven evidence that short-timescale opinion change in large online populations is effectively Markovian and governed by identifiable deterministic and stochastic forces, in a strongly polarized opinion landscape such as climate change. The proposed D-MODD framework captures both the systematic pull toward stable opinion attractors and the state-dependent variability that shapes how individuals transiently deviate from them. This establishes a rigorous generative model linking user-level behavioral trajectories to a compact macroscopic description of opinion dynamics at the level of an effective stochastic operator.

Importantly, the D–MODD framework is not restricted to stationary opinion landscapes. In principle, it can be extended to non stationary settings by allowing the drift and diffusion operators to vary in time,
$A(x,t)$ and $D(x,t)$, and by recalibrating them over sliding temporal windows.

From a sociological perspective, the inferred drift and diffusion functions provide a quantitative characterization of heterogeneity in individual opinion dynamics across the population, which can be interpreted as reflecting different degrees of opinion rigidity across the two communities. This interpretation is supported by the analysis of user profiles presented above, which reveals systematic differences between the two polarized groups. In particular, we observe that the denialist group is associated with trajectories of lower variability, indicating more stable and persistent opinion dynamics. This asymmetry in variability may contribute to the persistence of polarization in the climate change debate, independently of external information signals. This highlights the importance of state-dependent variability in shaping collective opinion patterns and provides an empirical link between individual-level behavioral heterogeneity and macroscopic polarization.

Beyond methodological novelty, our results show that opinion evolution in digital environments does not unfold as unstructured noise or purely through social imitation, but instead follows a coherent diffusion-like process with distinct attractor basins and spatially heterogeneous noise. This enables a quantitative characterization of online opinion dynamics in which theoretical models can be directly calibrated, tested, and refined against behavioral data.

More broadly, our approach demonstrates how the microstructure of social dynamics, often accessible only through stylized modeling assumptions, can be empirically recovered and represented within an effective stochastic framework, providing a foundation for predictive modeling of information flows, polarization, and collective behavior in large-scale social systems.

\section*{Data availability}

The anonymized opinion trajectories and interaction graphs used in this study will be made publicly available at \url{https://github.com/IxandraAchitouv/DMODD} upon publication along with the codes used to generate the figures.

\section*{Acknowledgement}
This research was partially supported by the Complex Systems Institute of Paris
Île-de-France (https://iscpif.fr) and the Paris Île-de-France program
SESAME 2021 - 00015415

\bibliography{reference}

\end{document}